\documentclass[aps,
              amsmath,amssymb,amsfonts,groupedaddress,
              onecolumn,
              showpacs,
              superscriptaddress,floatfix]{revtex4-1}
\usepackage[english]{babel}
\usepackage[pdftex]{color, graphicx}
\usepackage{epstopdf} 
\usepackage{subfig}
\usepackage{booktabs}
\usepackage{caption}
\begin{document}

\title{Coupling between whistler waves and slow-mode solitary waves}

\author{\sf {A. Tenerani}}
\affiliation{LPP, CNRS/Ecole Polytechnique/UPMC, St. Maur-des-Foss\'es, France }
\email{anna.tenerani@lpp.polytechnique.fr}
\affiliation{Physics Dept., University of Pisa, Pisa, Italy}

\author{\sf{F. Califano}}
\affiliation{Physics Dept., University of Pisa, Pisa, Italy}

\author{\sf{F. Pegoraro}}
\affiliation{Physics Dept., University of Pisa, Pisa, Italy}

\author{\sf{O. Le Contel}}
\affiliation{LPP, CNRS/Ecole Polytechnique/UPMC, St. Maur-des-Foss\'es, France }

%
%

\begin{abstract}
The interplay between electron-scale and ion-scale phenomena is of general interest for both laboratory and space plasma physics. In this paper we investigate  the linear coupling between whistler waves and slow magnetosonic solitons through two-fluid numerical simulations. Whistler waves can be trapped in the presence of inhomogeneous external fields such  as a density hump or hole  where they can propagate for times much longer than their characteristic time scale, as  shown by laboratory experiments and space measurements. Space measurements have  detected    whistler waves also in correspondence to magnetic holes, i.e., to density humps with magnetic field minima extending on ion-scales. This raises the interesting question of  how ion-scale structures  can couple to whistler waves. Slow magnetosonic solitons  share some of the  main features of a magnetic hole. Using the ducting properties of an inhomogeneous plasma  as a guide,  we present a numerical study of whistler waves  that are trapped   and transported inside propagating  slow magnetosonic solitons. 
\end{abstract}


\maketitle

\section{Introduction and aims}

The study of the interplay between phenomena at ion and at electron dynamical scales is crucial in order to understand the physical  mechanisms of  basic plasma processes such as energy dissipation and particle acceleration and heating in space and laboratory plasmas. Presently multi-point satellite measurements such as the Cluster satellites,  by combining space and time measurements, make it possible to investigate simultaneously electron-scale and ion-scale phenomena and to inspect  stationary  and propagating  magnetic and density structures. 

An important example of  coupling between ion-scales and  electron-scales  is provided by  the correlation   that has been found in space between  whistler waves, which  occur on the electron-scales, and magnetic field depressions associated to density humps that  have typical scale lengths of the order of the ion-scales, usually interpreted as non-propagating mirror mode structures~\cite{Smith_JGR_1976, Thorne_nature_1981, Tsurutani_JGR_1982, Baumjohann_An_Geo_1999, Dubinin_AG_2007}. On the other hand, whistler waves can also interact with slowly propagating Magnetohydrodynamic structures  involving  both density and magnetic field  modulations.  If the whistler waves become  trapped inside such  structures,  the problem arises how  low frequency nonlinear modes can act as carriers for  higher frequency waves.  

Whistler waves are electromagnetic right-handed polarized waves that propagate nearly parallel to the ambient magnetic field at frequencies in the range between the ion and the electron cyclotron frequency~\cite{Stenzel_JGR_1999}. A known property of whistlers is that in the presence of plasma inhomogeneities, such as magnetic field aligned tubes of density enhancements or depletions, their energy can be guided  for long times without being dispersed~\cite{Smith_JGR_1960, karpman_1981b, Karpman_Journal_Pl_Phys_1982, Streltsov_JGR_2006}. Examples of such ducted propagation has been found in satellite observations in the Earth's magnetosphere  \cite{Angerami_JGR_1970, koons_JGR_1989, Moullard_GRL_2002} and in laboratory plasmas \cite{stenzel_GRL_1976}.
\\
Magnetohydrodynamic waves also have been studied for decades  both in space and laboratory plasmas. Recently, growing attention has been devoted to the study of oblique slow magnetosonic solitary waves, namely  of coherent structures at the ion-scale characterized by a density hump and a magnetic field depression~\cite{McKenzie_PhysPl_2002, Stasiewicz_PRL_2003, Stasiewicz_PRL_2004, Stasiewicz_JGR_2005}. \\ 
Despite many studies of both magnetosonic and whistler waves, their coupling has not been investigated in detail and is yet poorly understood. In this paper we investigate the interaction between slow magnetosonic solitons and whistler waves by using a two-fluid model. Using the ducting properties of an inhomogeneous plasma as a guide, we propose a new mechanism of ducting and transport of whistler waves  arising from  a linear coupling with slow type magnetosonic solitons.  Such a mechanism could explain spacecraft observations in the Earth's magnetosphere, namely the recurrent detection of whistler waves correlated to magnetic holes. 

This  paper is organized as follows: in Section II we present  the two main  problems of interest separately: the whistler waves trapping and the slow mode solitary wave. In Section II A we treat the trapping of whistlers by an inhomogeneous magnetized plasma. The aim is to shed light on the ducting process using an equilibrium configuration simpler than the soliton solution. In this way we can obtain quantitative conditions to be used to estimate the ducting conditions  when the inhomogeneity is provided by the  magnetosonic soliton. In Section II B we introduce the slow mode solitary wave solutions of the two-fluid   equations and   discuss the role that these configurations  can have in trapping whistler waves. in Section III we describe  the system of equations of the two-fluid model used in our simulations, the initial conditions and the simulation  parameters. In Section IV we investigate  the trapping of whistler waves  by slow mode solitons  numerically and present the results. Finally, conclusions are discussed in Section V.

%
%

\section{Theoretical background}
\label{theo_back}

\subsection{Whistler wave trapping in magnetic and density ducts}
\label{whistler_trapping}

It has been shown \cite{Smith_JGR_1960, karpman_1981b, Karpman_Journal_Pl_Phys_1982, Streltsov_JGR_2006} that whistler waves propagating in a cold magnetized plasma in the presence of  inhomogeneities axially symmetric and transverse to the ambient magnetic field of the plasma density can be channelled by these  inhomogeneities (\emph{density duct}). The variation of the index of refraction caused by the density inhomogeneity makes the wave trajectory bend such that the average propagation is along the duct direction. As a consequence, the wave becomes  trapped by the duct.   

In this Section we extend these results  to the case where   both the equilibrium plasma density and magnetic field are inhomogeneous in the plane perpendicular to the direction of the magnetic field.

The aim of this analysis is to study the problem of whistler wave trapping with a \textquotedblleft simple\textquotedblright\ model. It will be used later as a reference for a more complex one, adopted in the numerical simulations, where the self consistent fields of the soliton are viewed by the whistler as perturbations of the ambient equilibrium. In order to mimic  the configuration of  interest (the slow mode soliton), we consider a plasma equilibrium characterized by a density hump and a magnetic field minimum, the \emph{magnetic hole}. 

\emph{Model --} We want to study the trapping properties of the equilibrium magnetic field inhomogeneities in the framework of a fluid model. Therefore we consider the regime $v_{th,e}\ll v_{ph}$ ($v_{th,e}$ and $v_{ph}$ being the electron thermal velocity and the whistler phase velocity respectively), and we  use the cold dielectric tensor of a magnetized plasma, $\varepsilon$. This  simplification  is convenient  because  the cold dielectric tensor $\varepsilon$ includes the basic effects of trapping thanks to its dependence on  the density and the magnetic field strength (through the electron cyclotron frequency $\omega_{ce}$). Assuming, for the sake of illustration, a two dimensional spatial configuration and taking the gradients along the magnetic field lines to be negligible on the scale of the whistler wave length, the plasma can be represented in a slab geometry, with density and magnetic field gradients perpendicular to the magnetic field direction. For the sake of clarity, we define \textquotedblleft magnetic hole\textquotedblright\ a fluid equilibrium with a density hump and a magnetic field depression perpendicular to the equilibrium magnetic field, as distinguished  from   a \textquotedblleft density duct\textquotedblright\  that  only  has a perpendicular density inhomogeneity. This model allows us to highlight the basic mechanism of the whistler wave trapping and to obtain  a quantitative estimate of the parameters to be used in the simulations, such as  the  value of the whistler frequency and the angle of propagation. We will show that a  magnetic hole requires less strict conditions on the whistler wave parameters than a density duct in order to trap whistler waves. 

\begin{figure}[h] 
\bigskip
   \centering
   \includegraphics[width=0.43\textwidth]{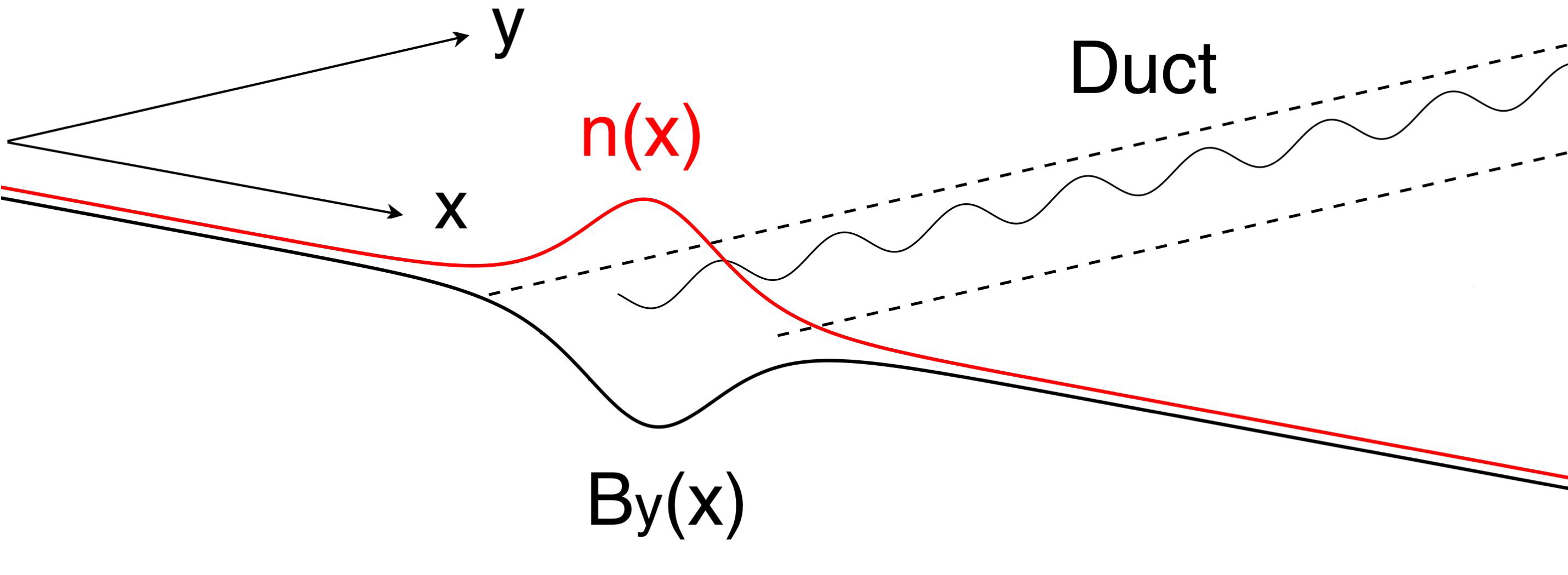} 
   \caption{\small{Schematic representation of the trapping of whistler waves: the wave propagates in the plane $(x,y)$ containing the magnetic field and the inhomogeneity direction and  is localized inside the density hump. In the case of a warm plasma, a magnetic field depression of the form $B_y(x)$ is associated to the density enhancement.}}
   \label{duct}
\end{figure}

Let us consider a bump-like density profile in the $x$ direction, perpendicular to the background magnetic field  ${\bf B}_y(x)$ which is taken to be directed along  $y$. The background magnetic field has  a minimum in correspondence to the density hump. A sketch of this configuration  is illustrated in Fig.~\ref{duct}: the wave propagates in the $(x,y)$ plane and  is localized inside the magnetic hole.
As shown in Ref.~\cite{karpman_1981b}, the set of the two-fluid equations for a magnetized plasma can be arranged so as to obtain the following system of equations for the electric and magnetic fields, where a time dependence of the form $\exp(-i\omega t)$ has been assumed:
\begin{equation}
\nabla(\nabla\cdot{\bf E})-\nabla^2{\bf E}=\frac{\omega^2}{c^2}(\varepsilon\cdot{\bf E})
\label{eq_onda_matrix}
\end{equation}
\begin{equation}
\nabla\cdot(\varepsilon\cdot{\bf E})=0
\label{poisson}
\end{equation}
\begin{equation}
{\bf B}=-i\frac{c}{\omega}\nabla\times{\bf E}.
\end{equation}
Looking for spatial solutions of the form $A(x)\exp(ik_{\parallel}y)$, where $k_{\parallel}\equiv k_y$ is the parallel wave vector, equation~(\ref{eq_onda_matrix}) can be reduced to two coupled differential equations of second order for the electric field components $E_x$ and $E_z$, while the third component $E_y$ is obtained from equation~(\ref{poisson}). By rescaling the variable $w=x/L$, where $L$ is the typical length of the large scale inhomogeneity such that $(c/\omega)/L\ll1$, the two coupled equations for $E_x$ and $E_z$ can be   solved by means of the WKB approximation \cite{karpman_1981b}. We impose solutions of the form
\begin{equation*}
A(w)=B(w)\exp\left(iL\omega/c\int^w{q(w')}dw'\right)
\end{equation*}
and retain only the terms to lowest order in $(c/\omega)/L$.  Note that such a  choice excludes the ducting mechanism described in \cite{Breizman_PRL_2000} where the whistler energy  propagates in the form of  \textquotedblleft surface waves\textquotedblright\ with frequency below the characteristic frequency of bulk whistler waves.  The  perpendicular wave vector $k_{\bot}\equiv(\omega/c)q=k_x$, for a given parallel wave vector $k_{\parallel}$ and frequency $\omega$, must satisfy the whistler dispersion relation of a homogeneous plasma  (in ion normalized units, see also Section~\ref{2fleq}) obtained in the limit $\omega_{ci}<\omega<\omega_{ce}\ll\omega_{pe}$:
\begin{multline}
k_{\bot,\,\pm}^2(x)=\frac{1}{2d_e^2\left({\omega}/{\omega_{ce}(x)}\right)^2}\times  \\
             \left\{ k_{\parallel}^2d_e^2\left[1-2\left(\frac{\omega}{\omega_{ce}(x)}\right)^2\right]-2n(x)  \left(\frac{\omega}{\omega_{ce}(x)}\right)^2\pm d_e k_{\parallel}\sqrt{d_e^2 k_{\parallel}^2-4n(x)\left(\frac{\omega}{\omega_{ce}(x)}\right)^2    }    \right\}.
\label{RD_q1,2}
\end{multline}
Note that, using ion normalized units, $d_e^2=(m_e/m_p)$ and $\omega_{ce}=Bm_p/m_e$. 

The whistler dispersion relation as expressed by equation (\ref{RD_q1,2}) for the complex variable  $k_{\perp}$  shows that, for fixed $k_{\parallel}$ and $\omega$, there are two \textquotedblleft branches\textquotedblright\ of the perpendicular wave vector corresponding to the plus and minus sign, respectively (the \textquotedblleft upper\textquotedblright and the \textquotedblleft lower\textquotedblright\ branch). For a solution given by a real $k_{\bot}$, we get a propagating whistler wave while for an imaginary $k_{\bot}$, we get a purely evanescent (non propagating) whistler wave.  The transition within a given branch from real to imaginary values is at the basis of the wave  trapping, and  is determined by the local values of the density $n$ and of the parameter $\omega/\omega_{ce}$.

The general WKB solution is given by a linear combination of the four wave solutions corresponding to the four possible wave vectors $\pm k_{\bot,\pm}$. Near the critical points, where $k^2_{\bot,\pm}=0$  or  $k^2_{\bot,-}=k^2_{\bot,+}$, the WKB approximation ceases to be valid. An analytical continuation of the solution in the complex $x$ plane around these points is therefore necessary in order to extend the solution to all its domain of validity. The complex  $x$ plane is divided into different portions by the so called Stokes and anti-Stokes lines which radiate out from the critical points \cite{heading}. When a Stokes line radiating from a turning point of a given branch is crossed, the two solutions of the given branch, say, the ones corresponding $+k_{\bot,-}$ and to $-k_{\bot,-}$, are coupled, corresponding to the reflection of a given wave when approaching a turning point. In an similar way, when a Stokes line radiating from a conversion point is crossed, the coupling between the two branches, for example $+k_{\bot,+}$ and $+k_{\bot,-}$, occurs. The coupling between the two branches leads,  for instance,  to the leakage of a propagating wave, again approaching a turning point. Because of the coupling with the other branch at the conversion point,  a fraction of the wave energy continues to propagate past the turning point~\cite{karpman_1981b}. The coefficients of reflection or conversion are exponentially small if the critical points are far from the real axis.

In this Section we consider only the trapping of whistler modes due to the presence of turning points on the real axis, where $k_{\bot,\pm}^2=0$. For simplicity the conversion between different branches will be neglected  here, assuming that the square root in equation~(\ref{RD_q1,2}) does not vanish on the real axis. Moreover, in order to look for solutions given by propagating waves or evanescent waves, we assume that $k_{\bot,\pm}^{2}$ is real, which means that the perpendicular wave vector is real (i.e. $k_{\bot,\pm}^{2}>0$) or imaginary (i.e. $k_{\bot,\pm}^{2}<0$). These assumptions correspond to the following condition:
\begin{equation}
k_{\parallel}^2>\frac{4n(x)}{d_e^2} \left(\frac{\omega}{\omega_{ce}(x)} \right)^2.
\label{condition_1}
\end{equation}

From equations~(\ref{RD_q1,2})--(\ref{condition_1}) it follows that the upper branch cannot be trapped in a magnetic hole (neither in a density duct). Indeed, $k_{\bot,+}^2$ is everywhere positive for frequencies $\omega/\omega_{ce}<1/2$, and thus the wave propagates in all regions. If $\omega/\omega_{ce}>1/2$ then $k_{\bot,+}^2$ is positive when $k_{\parallel}^2d_e^2>{n}/[{\omega_{ce}/\omega-1}]$. If this condition is satisfied inside the magnetic hole, it is satisfied  outside the magnetic hole as well, since $n(x)$ has lower values outside than inside the magnetic hole, and vice versa for the function $\omega_{ce}(x)$. Then also in this range of frequencies the wave propagates in all regions. We can therefore focus only on the lower branch $k_{\bot,-}$.

With the same reasoning as above, we see  that the lower branch can be trapped in a magnetic hole (or in a density duct) only for frequencies $\omega/\omega_{ce}<1/2$. Indeed, for frequencies $\omega/\omega_{ce}>1/2$ the perpendicular wave vector corresponding to the lower branch is imaginary everywhere while for  $\omega/\omega_{ce}<1/2$, the perpendicular wave vector is real  when 
\begin{equation}
k_{\parallel}^2< \frac{n/d_e^2}{\omega_{ce}/\omega-1},
\label{condition_2}
\end{equation} 
while it is imaginary when
\begin{equation}
k_{\parallel}^2> \frac{n/d_e^2}{\omega_{ce}/\omega-1}.
\label{condition_3}
\end{equation} 
To summarize, for frequencies $\omega/\omega_{ce}<1/2$,  trapping is possible in a magnetic hole (and in a density duct) if the parallel wave vector satisfies equation~(\ref{condition_1}) everywhere (which means that $k_{\bot}$ is either imaginary or real), and safisfies equation~(\ref{condition_2})  inside the magnetic hole, giving a propagating wave, and equation~(\ref{condition_3}) outside the magnetic hole, giving an evanescent wave. Since we are interested in whistler modes trapped into the magnetic hole,  from now on we consider only the lower branch in the frequency range $\omega/\omega_{ce}<1/2$ and drop the subscript \textquotedblleft$-$\textquotedblright. For the sake of clarity we define $n_{in}$ and $n_{out}$ the density calculated at the center of the magnetic hole (thus in correspondence to the minimum of the magnetic field and to the density maximum) and outside the magnetic hole (where the medium is homogeneous), respectively. By analogy we define the frequencies $\omega_{ce}^{in}$ and $\omega_{ce}^{out}$, the electron cyclotron frequency calculated at the center and outside the magnetic hole, respectively.  In this way, the trapping condition for the lower branch in a magnetic hole can be written as follows:
\begin{equation}
k_{inf}<k_{\parallel}<k_{sup},
\label{trapp_cond_max}
\end{equation}
where
\begin{equation}
k_{inf}(\omega)=max \left\{  \left[\frac{4n_{in}}{d^2_e}\,  \left(\frac{\omega}{\omega_{ce}^{in}}\right)^2 \right]^{1/2}\, , \left[ \frac{n_{out}}{d^2_e}\,\frac{1}{{\omega_{ce}^{out}}/{\omega}-1}\right] ^{1/2}  \right\}
\end{equation}
and
\begin{equation}
k_{sup}(\omega)=\sqrt{\frac{n_{in}}{d_e^2}\frac{1}{\omega_{ce}^{in}/\omega-1}}.
\end{equation}

\begin{figure}[h] 
   \centering
   \includegraphics[width=0.33\textwidth]{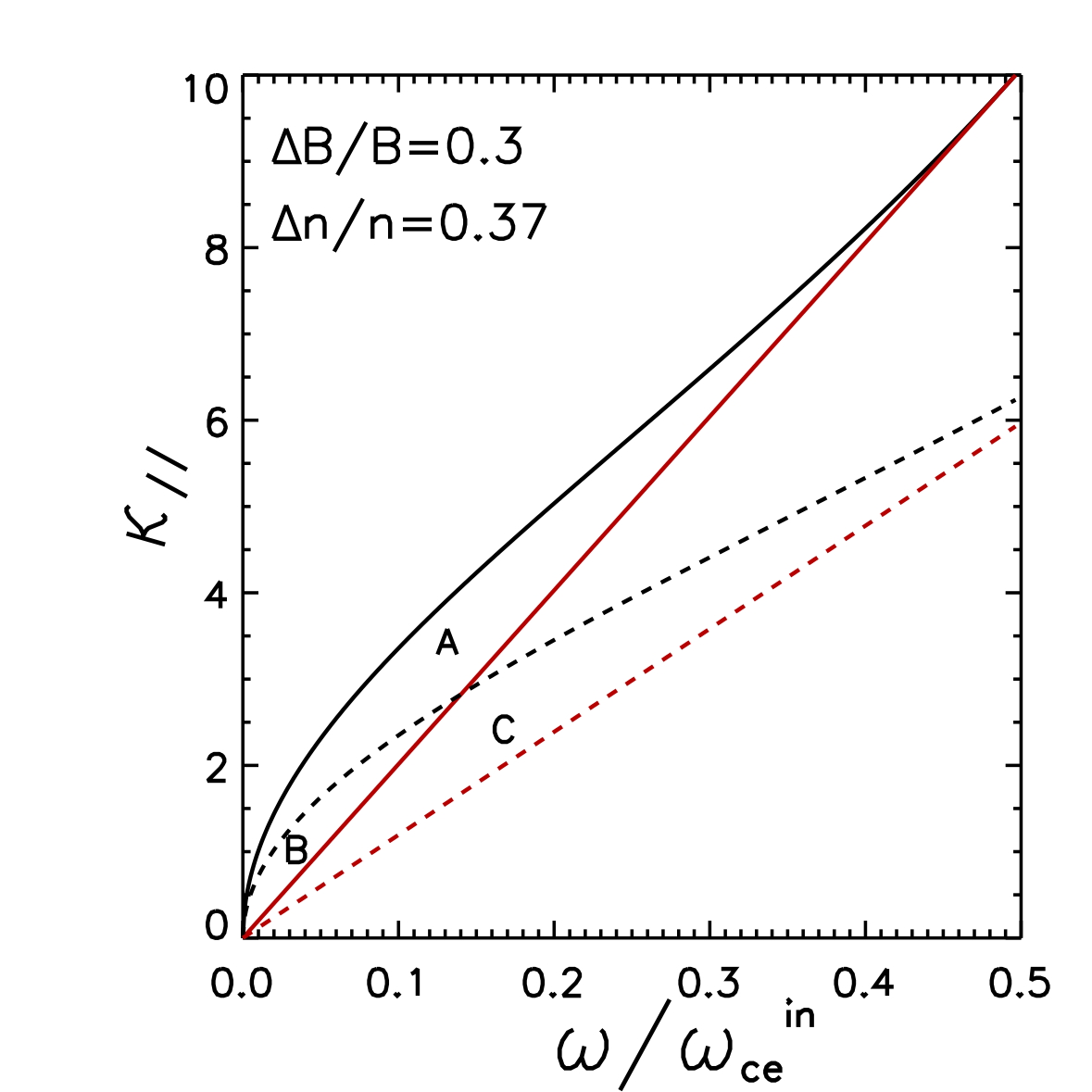} 
   \includegraphics[width=0.33\textwidth]{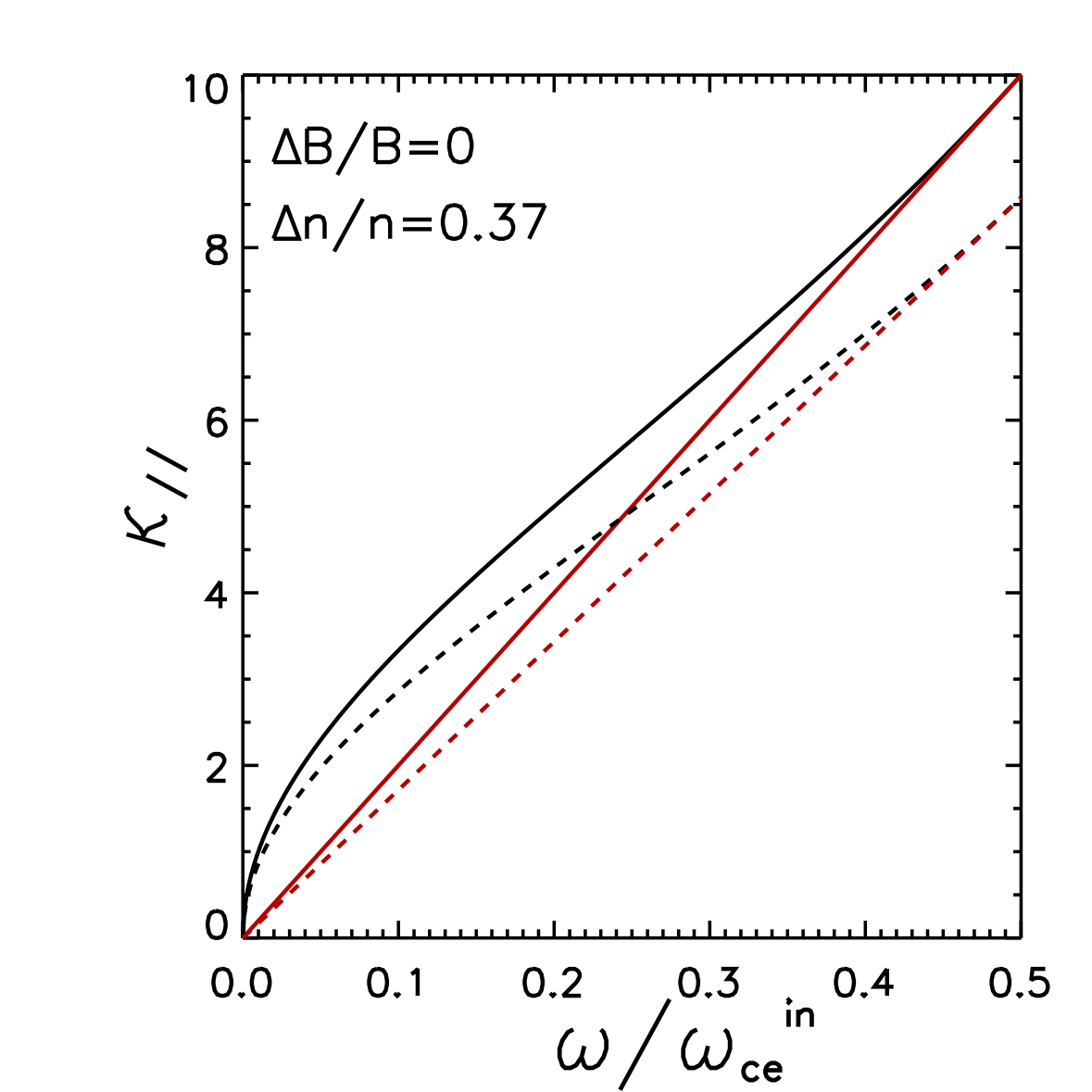} 
   \caption{\small{Plot of the curves which bound the region in the parameter space ($\omega/\omega_{ce}^{in},k_{\parallel}$) corresponding to real values of $k_{\bot,-}$. Solid and dashed lines are calculated in correspondence to the region inside and outside the channel, respectively. Red (straight) and black (curved) lines  correspond to the right-hand-side of equation~(\ref{condition_1})  and equation~(\ref{condition_2}), respectively. The portion {\sc a+b} and {\sc b+c} correspond to real values of $k_{\bot}$ inside and outside the inhomogeneous region, respectively. The portion {\sc a} corresponds to the trapped modes.  Left panel: plasma equilibrium with a magnetic hole; right panel: plasma equilibrium with a density hump on a uniform magnetic field.}}
   \label{k_lim}
\end{figure}

In Fig.~\ref{k_lim} we show a graphical representation of the portions in the parameter space ($\omega/\omega_{ce}^{in},k_{\parallel}$) corresponding to real values of $k_{\bot}$, calculated at the center (solid lines) and outside (dashed lines) the channel provided by the magnetic hole or the density duct. Red and black lines  correspond to the right-hand-side of equation~(\ref{condition_1})  and equation~(\ref{condition_2}), respectively. The left panel corresponds to a plasma equilibrium with a magnetic hole ($\Delta B/ B= |B^{in}-B^{out}|/B^{out}=0.3$ and $\Delta n/n = |n_{in}-n_{out}|/n_{out}=0.37$) and the right panel to a density duct with the same  density inhomogeneity than the magnetic hole ($\Delta B/ B=0$ and $\Delta n/n=0.37$). Referring to Fig.~\ref{k_lim}, left panel, the points ($\omega/\omega_{ce}^{in},k_{\parallel}$) lying in the portion {\sc a+b} and {\sc b+c} correspond to a propagating wave in the region inside and outside the channel, respectively. The intersection {\sc b} of these two regions corresponds to the untrapped modes, as they propagate both inside and outside the channel. The trapped modes are those corresponding to the portion {\sc a}, where $k_{\bot}$ is real inside and imaginary outside the channel. The maximum angle $\theta_{max}(\omega)$ of trapped modes for a given frequency is determined by  $k_{inf}$, and by the corresponding $k_{\bot}(\omega,k_{inf})$:
\begin{equation}
\theta_{max}(\omega)=\arctan \left[\frac{k_{\bot}(\omega,k_{inf})}{k_{inf}}\right].
\label{theta_max}
\end{equation}

A comparison between the magnetic hole, left panel in Fig.~\ref{k_lim}, and the density duct, right panel in Fig.~\ref{k_lim}, shows that the presence of magnetic variations (magnetic hole) leads to less strict trapping conditions. Indeed, for an equal density variation, the portion of trapped modes in a channel provided by both density and magnetic inhomogeneities is larger than in a channel formed only by a density inhomogeneity. In addition, the maximum angle of trapping (not shown here) results to be higher.

\subsection{Slow magnetosonic solitons acting as carriers of whistler wave energy}
\label{solitone}

We are interested in slow mode solitary waves because, as it will be explained in the following, they  can  trap whistler waves, similarly to magnetic holes.  In addition,  since solitary waves propagate almost unchanged, they provide an efficient channel that not only confines but also transports whistler energy at the typical soliton propagation speed.
 
Magnetosonic solitons are nonlinear one dimensional perturbations propagating in a warm plasma, obliquely to the equilibrium magnetic field \cite{Ohsawa_Phys_Fluids_1986, McKenzie_PhysPl_2002, Stasiewicz_JGR_2005}. Solitons are characterized by magnetic field and density perturbations in phase (fast solitons) or in opposition of phase (slow solitons). Solitary waves propagate with a constant profile and arise when the non linear terms are balanced by the dispersion terms. In a two-fluid model the required dispersion which gives rise to magnetosonic solitons is given by the Hall term and the electron inertia. Nonetheless, for non perpendicular propagations, the Hall term dominates the dispersion and the typical scales of solitons are $\sim d_i$. It can be shown that at some level of approximation the system of two-fluid equations can be reduced to a Kortveg de Vries equation~\cite{Ohsawa_Phys_Fluids_1986}, which has solitary wave solutions. 

Let us consider a soliton moving in the positive $x$ direction in a homogeneous magnetized plasma at rest, with equilibrium quantities defined as follows: 
\begin{equation*}{\bf B} = {\bf B}_0 = (B_{0x},\, B_{0y},\,0 ) \qquad {\bf u}_{i,\,e}=(0,\,0,\,0)\end{equation*}
\begin{equation*}n=n_{0} \qquad P_{i,\,e}=P_0,
\end{equation*}
where ${\bf u}_{i,\,e}$ and $P_{i,\,e}$ are the ion and electron velocity and pressure, respectively, $n$ the density and ${\bf B}$ the magnetic field. The angle of propagation of the soliton is defined as the angle between the direction of propagation, namely the $x$ direction, and the equilibrium magnetic field ${\bf B}_0$. For future convenience, in order to define the soliton direction of propagation, we will use the angle $\varphi_0$, taken as the complementary angle to the angle of propagation, thus defined as $\varphi_0=\arctan({B_{0x}/B_{0y}})$. An explicit solitary solution of the two-fluid system of equations is obtained in Ref.~\cite{Ohsawa_Phys_Fluids_1986} in the weakly non linear approximation. Both fast and slow solitons are found. Here we consider only the slow mode, as this mode has density and magnetic field in opposition of phase. The slow mode solitary solution depends on the amplitude of the perturbation $A$, on the angle  $\varphi_0$, on the temperature and on  the Alfv\'en speed which, in the homogeneous region, is equal to one in our units. The effects of the temperature enter through the sound speed $c_s$ and the slow magnetosonic phase speed $v_{p0}$, and  determines the velocity of propagation of the soliton and its width. Below, equations~(\ref{n_b_solit})--(\ref{n_solit_bis}) represent the total magnetic field and the total density of the plasma, defined as the values of the homogeneous equilibrium plus the fluctuations associated to the soliton, ${\bf B}_{tot}={\bf B}_0+{\bf B}_{sol}$, $n_{tot}=n_{0}+n_{sol}$: 
\begin{equation}
n_{tot} = 1 +  n_{sol}, \qquad B_{x,\,tot} = \sin{\varphi_0}, \qquad B_{y,\,tot} = \cos{\varphi_0} + \left[  \frac{(v^2_{p0} - c^2_s) }{\cos{\varphi_0}} \right]n_{sol},
\label{n_b_solit} 
\end{equation}
where 
\begin{equation}
n_{sol} \propto \frac{A}{ \cosh^2   \left[ \sqrt{\frac{A}{12\mu(\varphi_0, P_0)}} \, x     \right] }.
\label{n_solit_bis}
\end{equation}
The other plasma quantities are given explicitly in Appendix~\ref{eq_init_solit}, including the expression for the function $\mu(\varphi_0,P_0)$. Quantities are normalized to asymptotic equilibrium values outside the soliton.  

According to this theoretical analysis, the propagation speed of the soliton is  $V_0=v_{p0}+A/3$ and the typical width is $\ell\sim2\sqrt{12\mu/ A}$. The function $\mu$, which determines the width of the soliton, is a growing function of the temperature, ranging from values smaller than, or of the order of, $d_i$ to values much greater than $d_i$. Note that for the slow mode $v_{p0}<c_s$. Thus from the last of equations~(\ref{n_b_solit}), which defines $B_y^{tot}$, it follows that the magnetic field perturbation is in opposition of phase with the density perturbation. The analytical solution for the slow soliton is valid as long as the propagation is not parallel ($\varphi_0=\pi/2$) in which case $\mu$ equals zero (if $c_s<1$) or infinity (if $c_s>1$)~\cite{Ohsawa_Phys_Fluids_1986}. 

To summarize, the main features of slow mode solitons are: they carry a density hump perturbation associated to a magnetic field depletion and propagate obliquely with respect to the background equilibrium magnetic field at speeds which are much smaller  than that of whistler waves (greater than unity). This inhomogeneous system, which can be represented by an oblique solitary perturbation moving in a homogeneous plasma at rest,  is more complicated than the magnetic hole discussed previously, which instead has purely perpendicular gradients with respect to the magnetic field. Nevertheless, as a first approximation, it is possible to consider  the soliton perturbation superposed to the background equilibrium as a local and instantaneous magnetic hole for whistlers that are injected inside the soliton.


\section{Model equations, initial conditions and parameters}
\label{2fleq}
In order to describe numerically the trapping and transport of whistlers by solitary waves in a magnetized plasma, we use a quasi neutral adiabatic two-fluid model. The set of equations for ions and electrons, with labels  $i$ and $e$ respectively, are normalized using as characteristic quantities the ion mass density $nm_i$, the Alfv\'en velocity $v_a = B/\sqrt{¸4\pi nm_i}$ and the collisionless ion skin depth $d_i = c/\omega_{pi}$. With this choice $d_e^2=m_e/m_i$ and $\omega_{ce}=B/d_e^2$. The two-fluid model equations, in dimensionless form, are:
%
%
%
%
\begin{equation}
\label{2fleqadim1}
{\partial n}/{\partial t}+\nabla\cdot(n\mathbf{U})=0
\end{equation}
\begin{equation}
\label{adiab}
{\partial \mathcal{S}_{e,i}}/{\partial t}+\nabla\cdot(\mathcal{S}_{e,i}\mathbf{u}_{e,i})=0,\quad \mathcal{S}_{e,i}=P_{e,i}n^{1-\Gamma}
\end{equation}
\begin{equation}
\label{2fleqadim2}
{\partial (n\mathbf{U})}/{\partial t}=-\nabla \cdot \left[ {n} (\mathbf{u}_{i}\mathbf{u}_{i}+d_{e}^{2}\mathbf{u}_{e}\mathbf{u}_{e})  + (  P_{e}+P_{i}+{B^{2}}/{2}-\mathbf{BB}  ) \right].
\end{equation}
\begin{equation}
\begin{split}
\label{2fleqadim3}
(1-d_{e}^{2}\nabla^{2})&\mathbf{E}= - \mathbf{u}_{e} \times\mathbf{B} - (1/n)\nabla P_e-\\ &d_{e}^{2}\{ \mathbf{u_{i}\times B} - (1/n)\nabla P_i+ ({1}/{n})\nabla\cdot\left[ n(\mathbf{u_{i}u_{i}-u_{e}u_{e}}) \right] \},
\end{split}
\end{equation}
\begin{equation}
\label{2fleqadim4}
\mathbf{u_{e}}=\mathbf{U}-{\mathbf{j}}/{n}\ ,\  \mathbf{u}_{i}=\mathbf{U}+{d_{e}^{2}} {\mathbf{j}}/{n}\ ,\ \mathbf{U}= {\mathbf{u}_{i}+d_{e}^{2}\mathbf{u}_{e}},
\end{equation}
\begin{equation}
\label{2fleqadim5}
\mathbf{\nabla\times E}=-{\partial\mathbf{B}}/{\partial t}\ ,\  \mathbf{j}=\mathbf{\nabla\times B}+{\bf \hat{z}}j_z^{ext}.
\end{equation}
%
%
%
In these equations ${\bf E}$ and ${\bf B}$ are the electric and magnetic field, ${\bf u}_{i,e}$ is the ion (or electron) velocity, ${\bf j}$ the current, $j_z^{ext}{\bf \hat{z}}$ is an external forcing current used to inject whistlers, $P_{i,e}$ the pressure of ions and electrons $\Gamma$ is the adiabatic index.

The dimensions of the simulation box, $L_x$ and $L_y$, and the resolution of the grid, $dx$ and $dy$, are chosen in order to find a compromise between the different time and length scales at play. $L_y$ is chosen in order to let the whistler wave train propagate for several tens of $\omega_{ci}^{-1}$, without reaching the boundaries. $L_x$ is chosen in order to contain the soliton which is wider or of the order of $d_i$ and moves at a speed $V\sim0.1$. Finally, the mesh size must resolve the whistler wavelength. In Table~\ref{whist_Table} we report the parameters of the simulation box. The ion to electron mass ratio is fixed to $m_p/m_e=100$.
\begin{table}[h]
\centering
\begin{tabular}{l  c c c c c c c  c c}
\hline
   & $\omega_0$ & $\omega_0/\omega_{ce}^{in}$ & $\theta$&  $L_x$&$L_y$ & $dx$& $dy$ &$\ell$\\
\hline
Sim.~1      & 2.37    & 0.03   & -0.198  &   $24\pi$  & $160\pi$  & 0.08     & 0.1 & $2$\\
Sim.~2      & 2.37  & 0.03   & 0.6         &    $24\pi$  & $60\pi$   & 0.08     & 0.04 & $2$\\
Sim.~3      & 2.37  & 0.03   & 1.25       &    $24\pi$  & $60\pi$   & 0.08     &  0.04& $2$\\
Sim.~4      & 8        & 0.1      & -0.198   &   $24\pi$  & $60\pi$   & 0.08     &  0.04& $2$\\
Sim.~5      & 8     &   0.1       & 0.6        &   $24\pi$  & $60\pi$   &  0.08    &  0.04& $2$\\
Sim.~6      &   8   &   0.1      &   1.3      &  $24\pi$  & $60\pi$   &  0.08    & 0.04& $2$ \\
Sim.~7      &   3   &   0.04      &   -0.24    &  $24\pi$&  $160\pi$   &  0.08   &  0.1& $13$ \\
Sim.~8      &   3   &   0.04      &   0.3      &  $24\pi$ &  $120\pi$  &  0.08   & 0.08 & $13$ \\
Sim.~9      &   3   &   0.04      &   0.6     &  $24\pi$  & $240\pi$  &  0.08   & 0.16 & $13$\\
\hline
\end{tabular}
\caption{\small{Parameters of the injected whistler mode, of the simulation box and the characteristic width of the soliton $\ell$. }}
\label{whist_Table}
\end{table}

\begin{figure}
\begin{center}
\setlength{\unitlength}{0.8cm}
\begin{picture}(0,6.5)
\multiput(-1.2,0)(0,1){5}{\line(0,1){0.3}}
\multiput(2.2,0)(0,1){5}{\line(0,1){0.3}}
\put(-3,0){\vector(1,0){6}}
\put(2.8,0.2){$x$}
\put(-3,0){\vector(0,1){6}}
\put(-2.8,5.7){$y$}
\put(0.5,0){\line(0,1){4}}
\put(0.5,0){\vector(1,1){1.4}}
\put(0.5,0){\vector(2,3){1.4}}
\put(-3,0){\vector(1,3){1.2}}
\put(1.22,0.9){\oval(0.35,0.35)[t,r]}
\put(0.5,0.85){\oval(1.12,1.24)[t,r]}
\put(0.5,0.9){\oval(1.2,1.2)[t,r]}
\put(-3,1.05){\oval(0.7,0.8)[t,r]}
\put(1.2,2.3){${\bf B}_{tot}^{in}$}
\put(-2.6,3.4){${\bf B}_{0}$}
\put(0.8,1.7){$\varphi$}
\put(-2.88,1.9){$\varphi_0$}
\put(1.7,0.8) {${\bf k}$}
\put(1.45,1.17){$\theta$}
\put(0.25,4.8){$\longrightarrow$}
\put(0.5,5.1){$V$}
\end{picture}
\end{center}
\caption{\small{Schematic view of the system. The dashed lines indicate the region filled by the soliton moving in the positive $x$ direction with velocity $V$; ${\bf B}_{tot}^{in}$ is the total magnetic field at the center of the soliton, forming an angle $\varphi$ with the $y$ axis, and ${\bf k}$ is the whistler wave vector, which forms an angle $\theta$ with the magnetic field ${\bf B}_{tot}^{in}$. Outside the soliton the equilibrium magnetic field ${\bf B}_0$ forms an angle $\varphi_0$ with the $y$ axis.}}
\label{schema}
\end{figure}
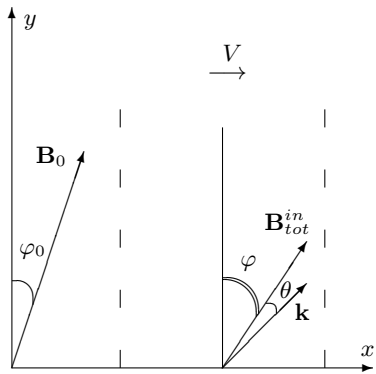

As initial condition, we consider a slow mode solitary wave centered in the simulation domain and superposed to a homogeneous magnetized plasma at rest. Oblique whistlers, as explained in the following,  are injected artificially in the simulation box, during the initial phase, in correspondence to the soliton. In order to do this we make use of an oscillating forcing current lasting over a characteristic time $\tau$.  In Fig.~\ref{schema} we show a schematic view of the system. The dashed lines indicate the region filled by the soliton moving in the positive $x$ direction with velocity $V$; ${\bf B}_{tot}^{in}$ is the total magnetic field at the center of the soliton, forming an angle $\varphi$ with the $y$ axis, and ${\bf k}$ is the whistler wave vector. In particular, when whistlers are generated inside the soliton, the subscripts  \textquotedblleft$\bot$\textquotedblright and \textquotedblleft$\parallel$\textquotedblright\ of the wave vector refer  to the total magnetic field ${\bf B}_{tot}^{in}$ at the center of the soliton.  Outside the soliton the total magnetic field reduces to the equilibrium magnetic field ${\bf B}_0$ forming an angle $\varphi_0$ with the $y$ axis. 

The analytical form of the fields representing the soliton superposed to the homogeneous equilibrium  at the initial time are given in Appendix~\ref{eq_init_solit}. The parameters of the soliton, listed in Table~\ref{solitons}, are chosen in order to have a narrow soliton with a width $\ell$ of the order of the ion skin depth (simulations 1 to 6) or a wider soliton with $\ell$ of the order of several ion skin depths (simulations 7 to 9).

To our knowledge, the numerical stability of the approximate solutions discussed in Section~\ref{solitone} has never been investigated. Therefore, before considering the full problem including the injection of the  whistlers, we tested the stability of the soliton solutions numerically. Our results show that they are well stable in the range of propagation angles $\varphi_0=0.57-0.17$ and of typical variations with respect to the equilibrium $n_{sol}\sim0.02-0.8$, $|B_{y,\,sol}|/B_{0y}\sim0.01-0.5$. They propagate at the expected velocity maintaining almost unchanged their initial profile over times $t\sim1000$, until they exit from the simulation box.

\begin{figure}[htbp]
\centering
\subfloat{
{\includegraphics[width=.33\textwidth]{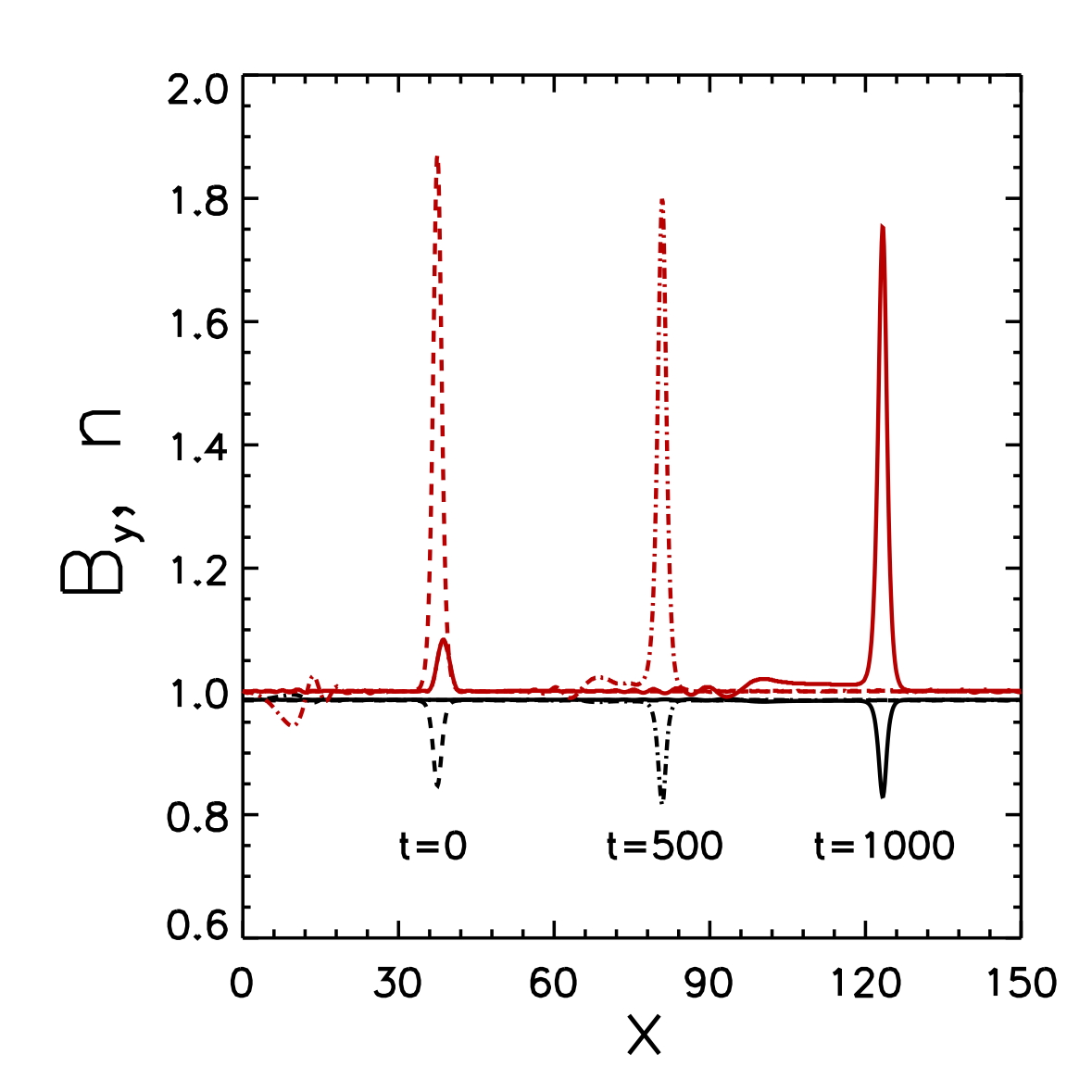}}\label{solit_1}} \quad
\subfloat{
{\includegraphics[width=.33\textwidth]{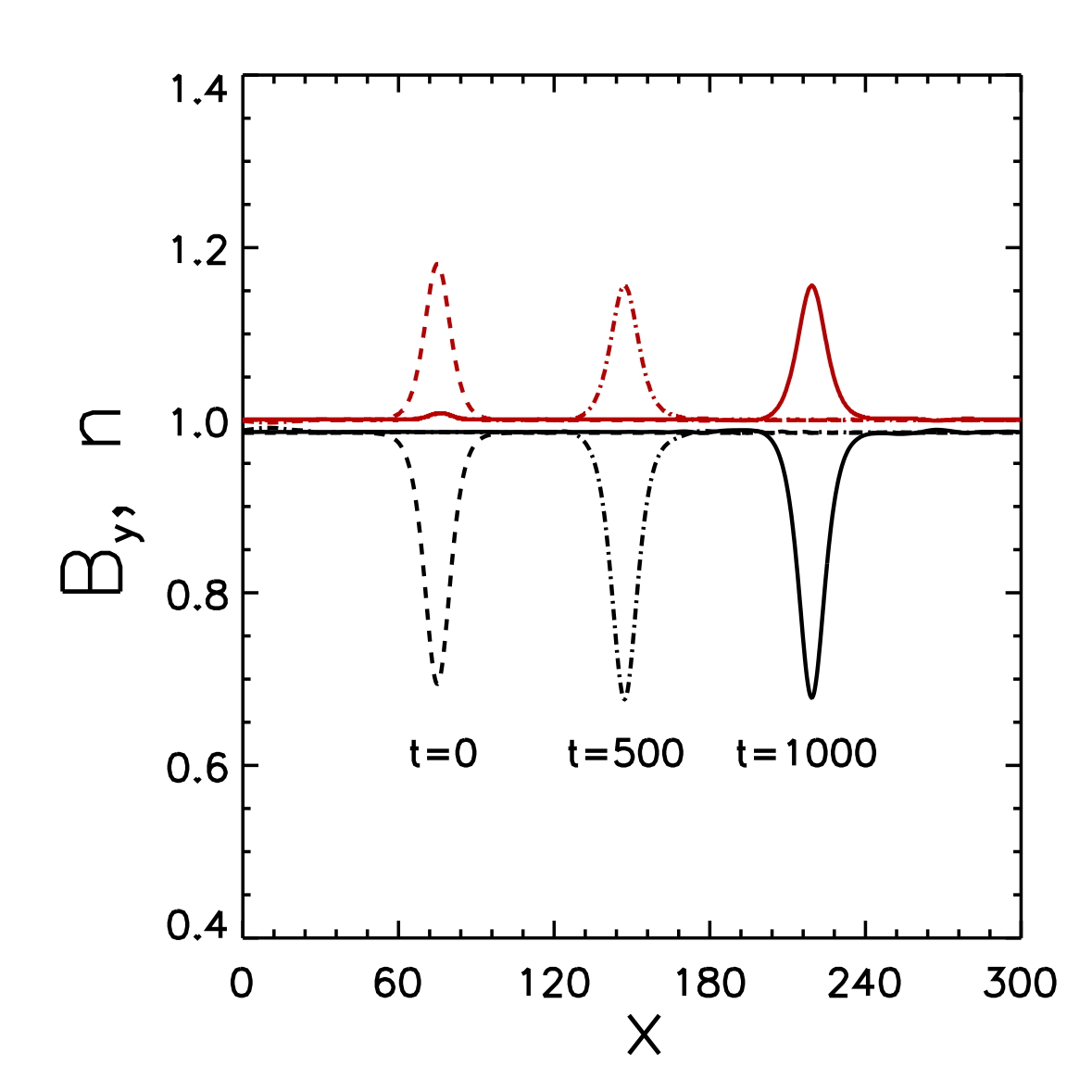}}\label{solit_2}}
\caption{\small{Examples of slow magnetosonic solitons at three different times: the black lines (depletions) represent the magnetic field $B_{y,\,tot}$ and the red lines (humps) the density $n_{tot}$. In the left panel we represent a narrow, strong amplitude soliton ($\Delta n/n\sim0.9$) and in the right panel a wider and weaker soliton ($\Delta n/n\sim0.2$). Notice that the initial soliton profile slightly modify during the temporal evolution, especially for large amplitude solitons, since the analytical profile implemented as initial condition is not an exact solution of the two-fluid system.}}
\label{solit}
\end{figure}

In Fig.~\ref{solit} we show two examples of quasi perpendicular ($\varphi_0=0.17$) magnetosonic solitons propagating along the $x$ axis at three different times up to $t\sim1000$. The red lines represent the density profile $n_{tot}$ and the black lines the magnetic field $B_{y,\,tot}$. In the left panel we represent a narrow, strong amplitude soliton ($\Delta n/n\sim0.9$) and in the right panel a wider and weaker soliton ($\Delta n/n\sim0.2$). Notice that the initial soliton profile slightly modify during the temporal evolution, especially for large amplitude solitons, since the analytical profile is not an exact solution of the two-fluid system. For the sake of clarity, we indicate with a subscript \textquotedblleft tot\textquotedblright\ the quantities resulting from the sum of the homogeneous background equilibrium plus the soliton perturbations. These large scale variation fields  can be considered as  the \textquotedblleft inhomogeneous equilibrium\textquotedblright\ supporting the whistler waves. 
\begin{table}[h]
\centering
\begin{tabular}{c c c c c c c c c c c}
\hline
&$A$ &  $P_{0\, i,e}$ &$\varphi_0$ & $\ell$ & $B_{y,\,tot}^{in}$ & $n_{tot}^{in}$& $\varphi$ & $V$& $\omega_{ce}^{in}$\\
\hline
Sim.~1--6&$0.073$  &  0.05 &0.17   &  $2$   & 0.85  & 1.87 & 0.2  & 0.1 & 86\\
Sim.~7--9&$0.034$  &  0.5   &0.17   &  $13$ & 0.7  & 1.18 & 0.24  & 0.1 & 72\\
\hline
\end{tabular}

\caption{\small{Parameters of the solitons.}}
\label{solitons}
\end{table}

Finally, we generate small amplitude oblique whistlers with frequency $\omega_0$ and propagating at an angle $\theta$ with respect to the \emph{total} magnetic field during a characteristic time scale $\tau$ by an external forcing corresponding to a current along the $z$ axis, $J_z^{ext}$ (see equation~(\ref{2fleqadim5})). The parameters of the forcing and thus of the injected whistlers, listed in Table~\ref{whist_Table}, are chosen making use of the trapping conditions discussed in Section~\ref{whistler_trapping}. In particular, we use as reference the maximum trapping angle defined by equation~(\ref{theta_max}). The maximum angle of trapping at a given frequency $\omega_0$, that we define as  $\theta_{max}^{\omega_0}$, allows us to  choose the frequency $\omega_0$ and the angle $\theta$ properly  in order to inject a specified whistler mode that we expect to be trapped or not. 
%
%
%
%
%
%
%
%
%

\section{Trapping of whistler waves by slow magnetosonic solitons: numerical  results}
\label{trapping_solit_whist}

In this section we show, by means of numerical simulations, that whistlers can be trapped and transported away by a slow magnetosonic soliton. Even if a slow soliton propagating in a homogeneous magnetized plasma is more complicated than the so called magnetic hole, as a first approximation the same properties  of whistler ducting apply, and the trapping conditions found for the magnetic hole discussed in Section~\ref{whistler_trapping} are therefore a good reference when asking which whistler modes can be trapped by the soliton.  

We have investigated the slow magnetosonic ducted and unducted regime of whistler modes by varying the typical width of the soliton. Here we report two different typical cases: a narrow soliton of width $\ell\sim 2\lesssim k^{-1}$ and a wider  soliton of width $\ell\sim 13> k^{-1}$ (in units of $d_i$), where $k$ is the whistler wave vector estimated for a given frequency and propagation angle from the two-fluid cold dispersion relation (see equation~(\ref{RD_w_2fluid}) in Appendix~\ref{two_fl_cold_RD}). A list of the parameters used in the simulations for the \textquotedblleft small\textquotedblright\ and  \textquotedblleft large\textquotedblright\ soliton are listed in Table~\ref{solitons}. The wider soliton has a weaker density hump but a stronger magnetic field depression than the narrow one. The injected whistler modes fluctuate at low frequencies ($\omega_0\ll0.1\,\omega_{ce}^{in}$) or high frequencies ($\omega_0\sim0.1\,\omega_{ce}^{in}$) with different angles of propagation ranging from $\theta\ll\theta_{max}^{\omega_0}$ to $\theta>\theta_{max}^{\omega_0}$. In the following, we focus on two simulations, namely Sim.~1 for the narrow soliton and Sim.~7 for the wide soliton, to show the trapping of whistlers.  

\begin{figure}[htbp]
\centering
\subfloat{
{\includegraphics[width=.29\textwidth]{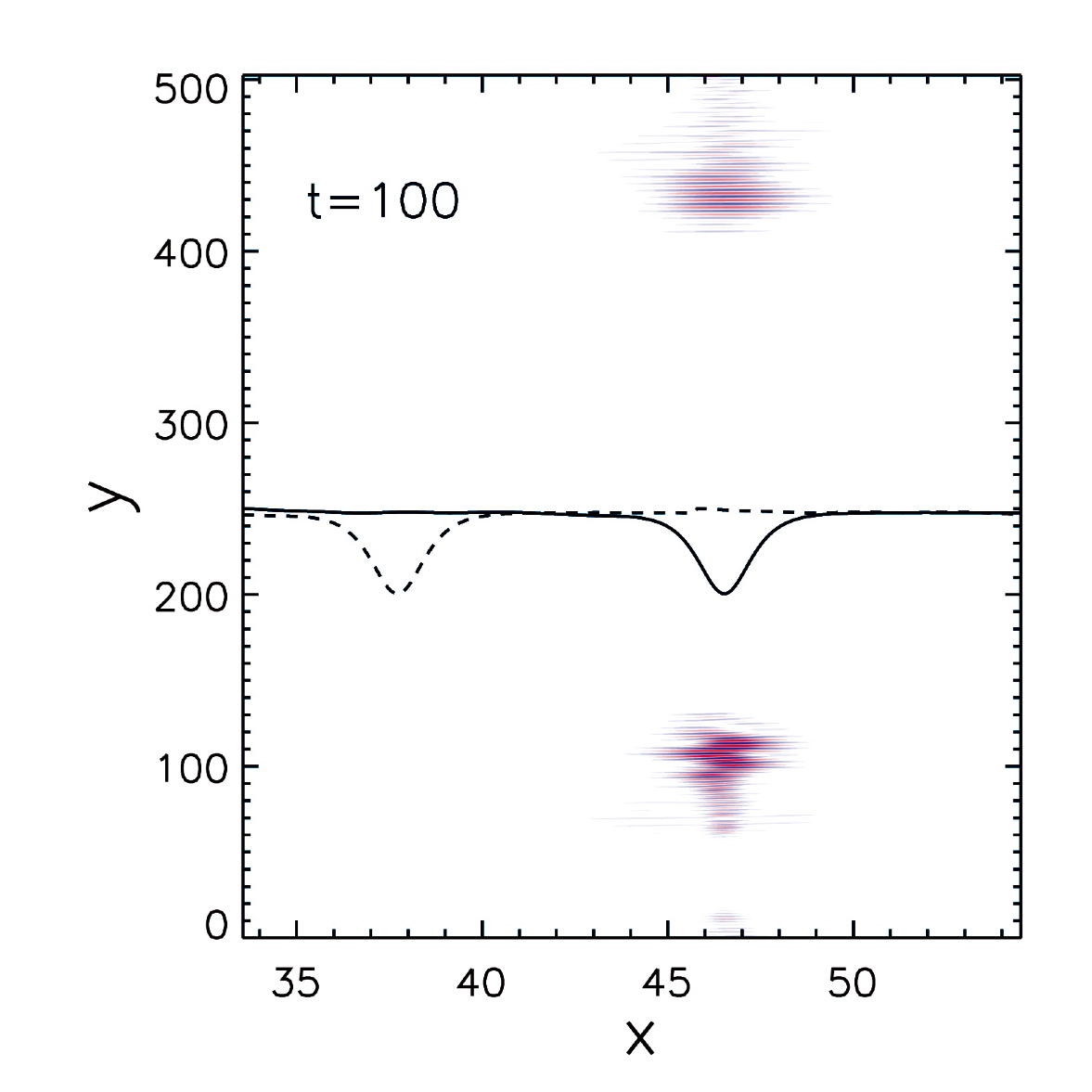}}
\label{sol1_sim1_t100}} \quad
\subfloat{
{\includegraphics[width=.29\textwidth]{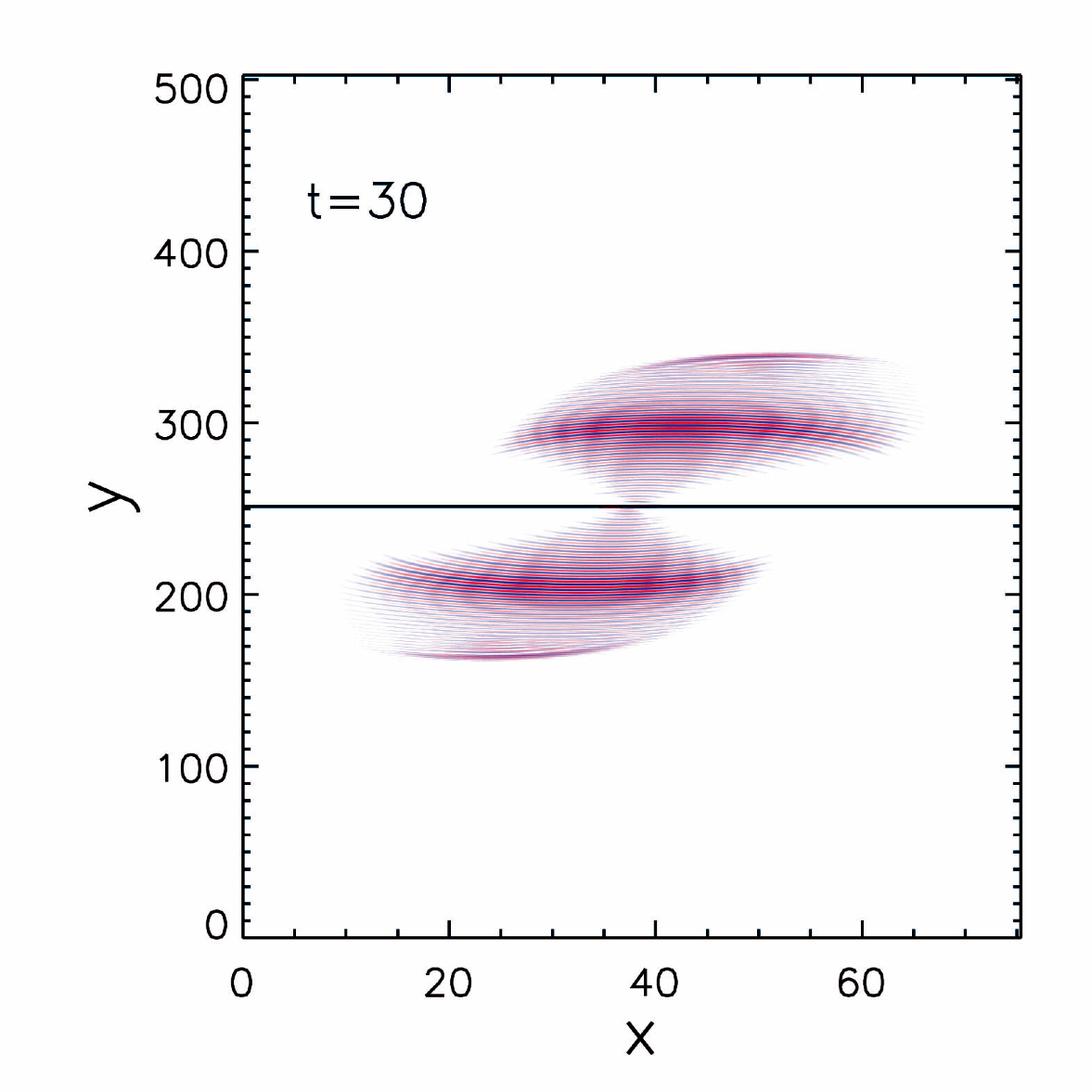}}
\label{sol1_sim1omo_t30}}\quad
\subfloat{
{\includegraphics[width=.29\textwidth]{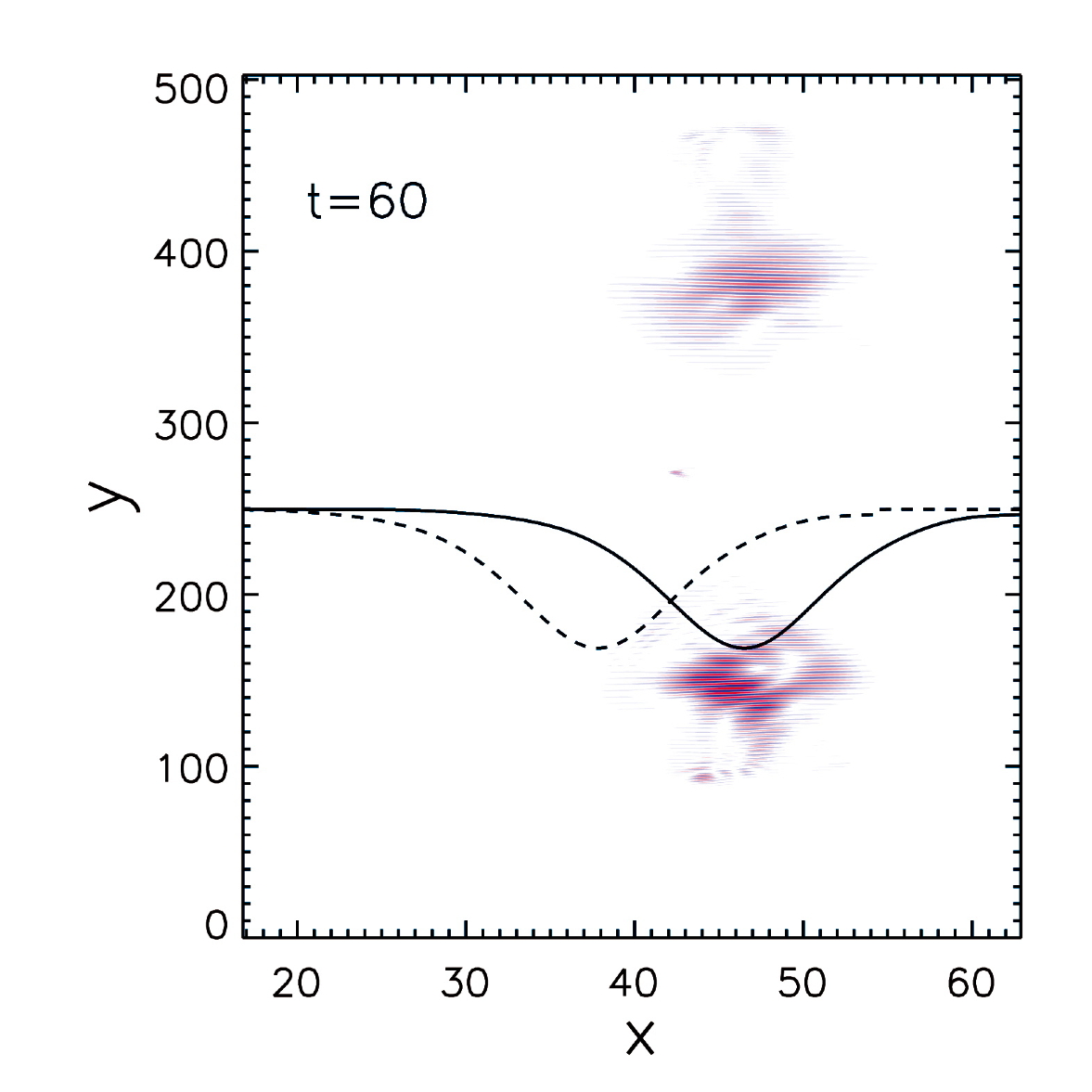}}}
\caption{\small{Contour plot of the $x$ component of the whistler magnetic field $b_x$. The profile of the soliton is represented by the black lines. The dashed line represents the soliton at time $t=0$. Left panel: trapping of whistlers in Sim.~1 at time $t=100$. Middle panel: the whistler wave packets in the entire space domain at $t=30$ as in Sim.~1 but in the absence of the soliton. Right panel: trapping of whistlers in Sim.~7 at time $t=60$.}}
\label{whist_sim}
\end{figure}

In these simulations the injected whistlers have frequency $\omega_0\sim0.03\,\omega_{ce}^{in}$ and $\omega_0\sim0.04\,\omega_{ce}^{in}$ respectively. They are injected along the $y$ axis, slightly oblique with respect to the  local total magnetic field, forming an angle $\theta=-0.198$ and $\theta=-0.24$, respectively, then satisfying $|\theta|\ll\theta_{max}^{\omega_0}$. The forcing current oscillates at the center of the simulation domain and switches off exponentially on a characteristic time shorter with respect to that of the soliton propagation. In this way, the forcing generates two finite size wave packets  in the $(x,y)$ plane propagating away from the source region in the two opposite directions, namely in the positive $y$ direction (upward) and in the negative $y$ direction (downward). The two wave packets propagate upward or downward, respectively, and remain spatially confined along the inhomogeneous $x$ direction in correspondence to the soliton, following its displacement along $x$. This is shown in Fig.~\ref{whist_sim} by the contour plots of the $x$ component of the magnetic field $b_x$ of the whistler waves in the simulation domain when the current has switched off and the wave packets are well developed. The profile of the soliton is represented (not in scale) by black lines and the dashed line corresponds to the soliton at time $t=0$. The left panel represents the two wave packets at time $t=100$, for the narrow soliton. In the middle panel we show the propagation of the same wave packets as injected in Sim.~1, but in a homogeneous equilibrium, i.e. with  $B_{0y}=B_{y,tot}^{in}=0.85$ and $n_0=n_{tot}^{in}=1.87$. We see that in the absence of the soliton the injected wave packets spread out during propagation. Finally, the right panel shows the wave packets at time $t=60$, for the wide soliton. To summarize, our simulations provide evidence that the waves, initially  injected inside the soliton structures, propagate along the duct provided by the soliton, upward or downward, advected at the same time in the perpendicular $x$ direction by the soliton. The whistlers are thus confined and transported by the slow soliton over times much larger than their typical time scale. 

\begin{figure}[htbp]
\centering
\subfloat{
{\includegraphics[width=.29\textwidth]{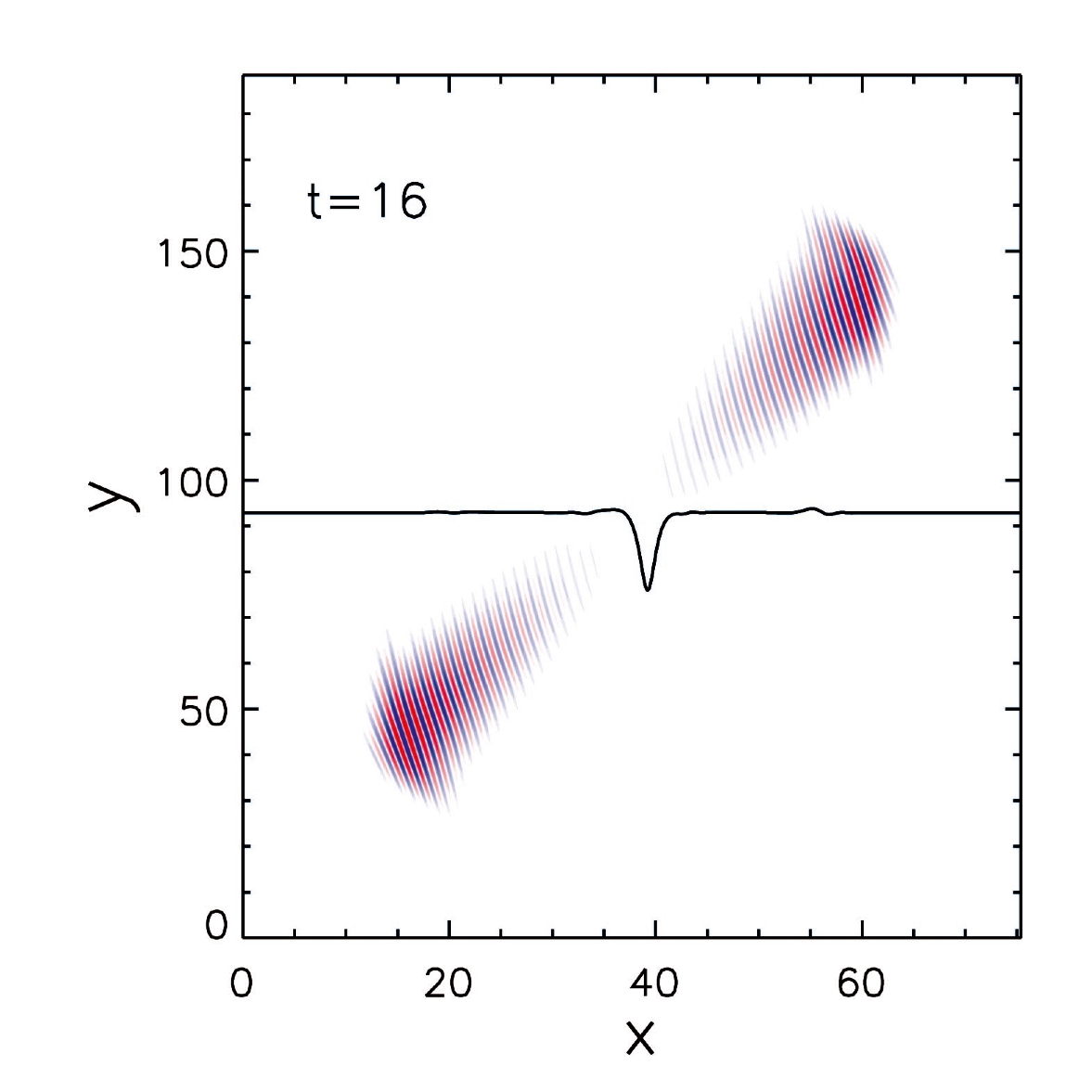}}
\label{sol1_sim6_t16}} \quad
\subfloat{
{\includegraphics[width=.29\textwidth]{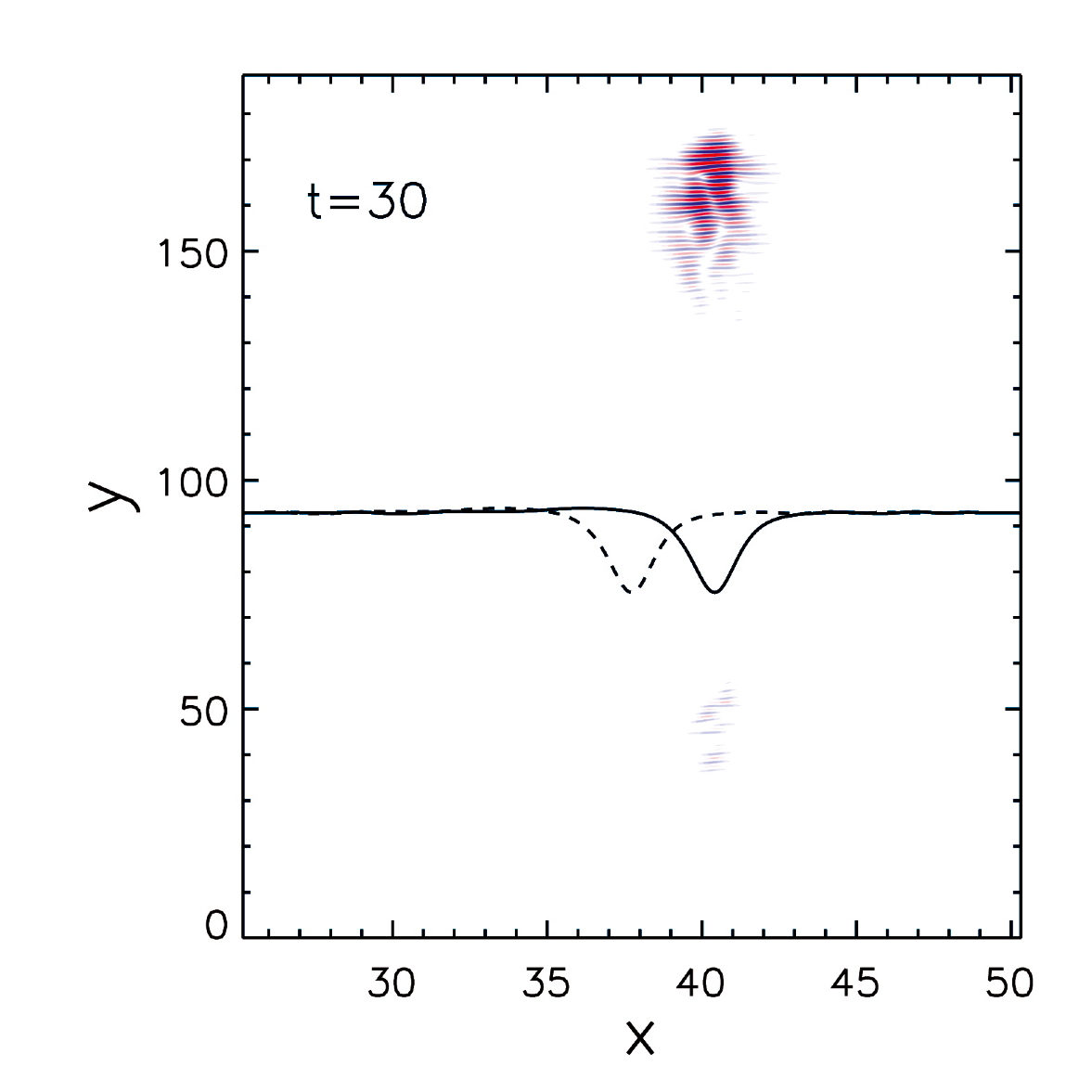}}
\label{sol1_sim4_t30}}\quad
\subfloat{
{\includegraphics[width=.29\textwidth]{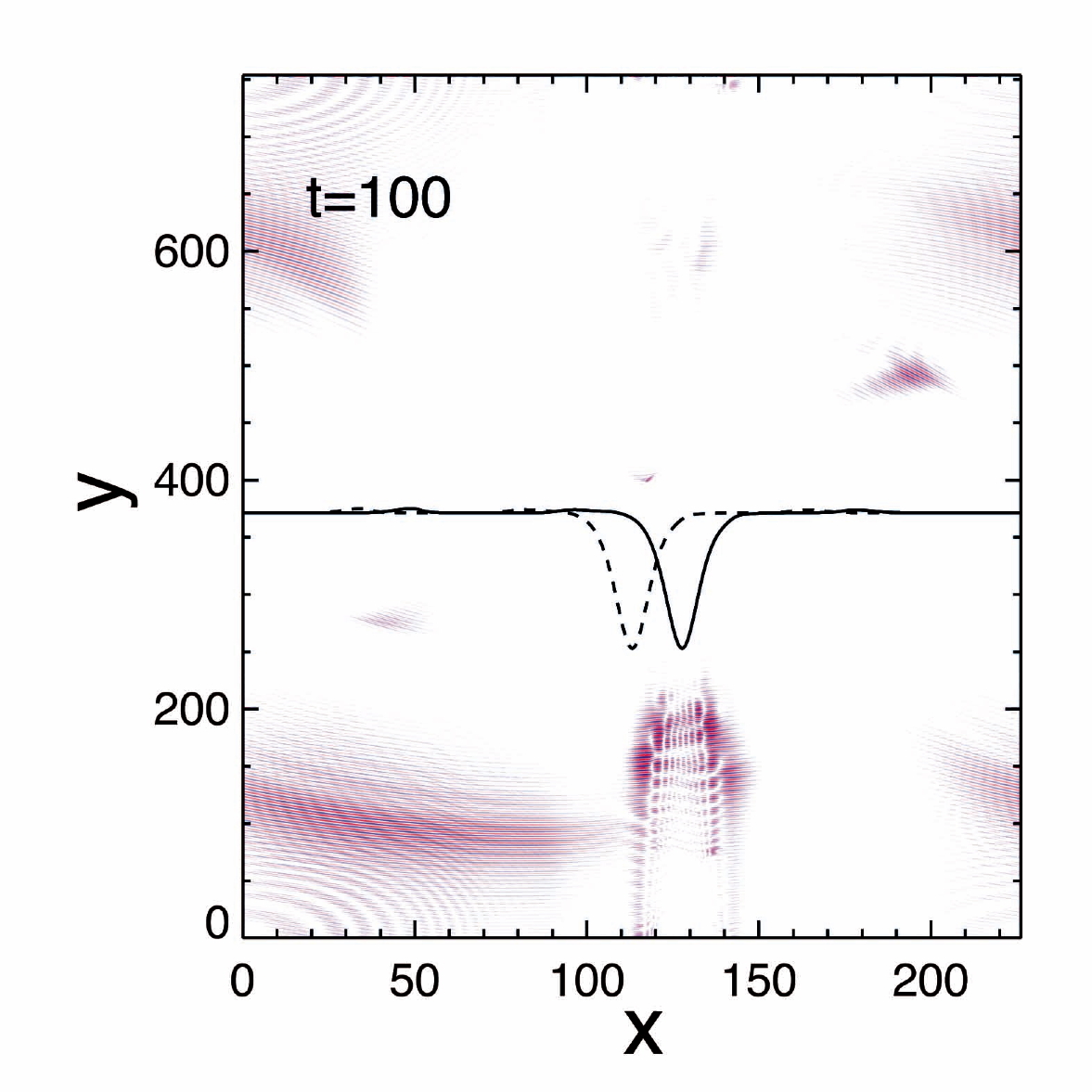}}
\label{sol2_sim9_t16}}
\caption{\small{Contour plot of the $x$ component of the whistler magnetic field $b_x$, with the profile of the soliton represented by the black lines. The dashed line corresponds to the soliton at $t=0$. Left panel: Sim.~6 at time $t=16$, corresponding to a high frequency, highly oblique whistler that escapes from the narrow soliton. Middle panel: Sim.~4 at time $t=30$, corresponding to a high frequency whistler that is trapped only in the upward direction while the downward wave packet becomes evanescent after one reflection. Right panel: Sim.~9 at time $t=60$, corresponding to an upward whistler that escapes outside the soliton while the downward wave packet is trapped. The periodic boundary conditions cause the waves approaching the upper (lower) boundary of the simulation box to appear in the lower (upper) boundary.}}
\label{whist_no_trap}
\end{figure}

We consider now two simulations, Sim.~3 and Sim.~6 where we inject  highly oblique whistlers with $\theta>\theta_{max}^{\omega_0}$. In this case the waves escape outside the solitons. An example is shown in Fig.~\ref{whist_no_trap}, left panel, where we show the contour of $b_x$ and the profile of the soliton at time $t=16$ in the case of Sim.~6. Here the injected whistlers are at high frequency ($\omega_0\sim0.1\,\omega_{ce}^{in}$) and at an angle of propagation $\theta=1.3$.

These results are in good agreement with the ducting theory. However, the model used in our numerical study is far richer than the reference model of the magnetic hole (Section~\ref{whistler_trapping}), and there are important effects that can modify the trapping conditions.

\begin{figure}[htbp]
\centering
\subfloat{
{\includegraphics[width=.35\textwidth]{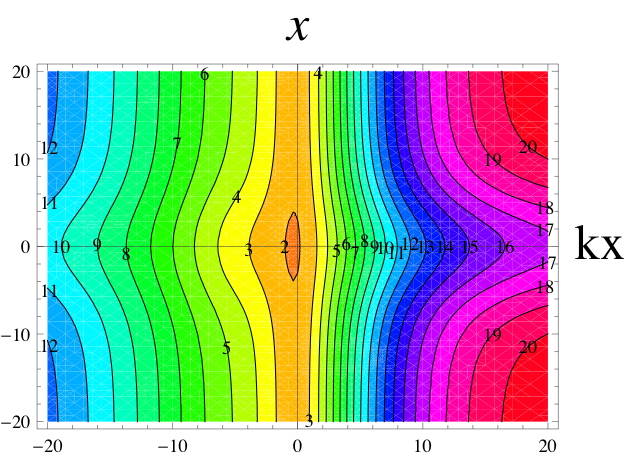}}
\label{sol_2_th_0_6U}} \quad
\subfloat{
{\includegraphics[width=.35\textwidth]{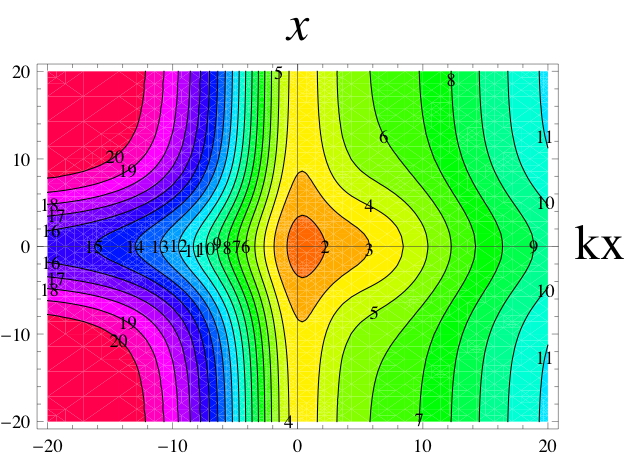}}
\label{sol_2_th_0_6D}}
\caption{\small{Contours of the whistler dispersion relation in the rest frame of the soliton, $\omega_0({\bf k}, x)=\omega({\bf k},x)+{\bf k\cdot U}(x)$, in the~$(k_x,x)$ plane with the parameters of Sim.~9. The left panel refers to the upward wave packet, which has $k_y\sim1.3$, and the right panel to the downward wave packet, which has $k_y\sim-1.7$. Whistler wave packets with a given $k_y$ evolve moving along the orbits at constant frequency. In Sim.~9 the frequency is $\omega_0=3$ that corresponds to an open orbit for the upper wave packet and to a closed orbit for the lower wave packet.}}
\label{num_mathematica}
\end{figure}

First of all, even if we neglect the displacement of the soliton, there is a finite perturbation in the plasma velocity of the form ${\bf U}(x)$ (see equations (\ref{Ux})--(\ref{Uz}) in Appendix \ref{eq_init_solit}). The presence of the fluid velocity introduces an asymmetry in the system, due to the Doppler effect, between wave packets propagating upward and downward, $k_y>0$ and $k_y<0$, respectively. Second, there are gradients along the total magnetic field, which can drive whistlers outside the soliton even if trapping conditions are satisfied. In the case of the magnetic hole $k_{\parallel}$ and $\omega$ are fixed quantities, while for the soliton $\omega$ and $k_y$ are constant but the parallel wave vector varies as the whistler propagates inside the soliton. As a consequence, while the whistler propagates towards the edge of the soliton, $k_{\parallel}$  can approach the value $k_{inf}$ (as defined in Section~\ref{whistler_trapping}, see also Fig.~\ref{k_lim}, left panel) thus allowing the whistler to become untrapped. An example is given in Fig.~\ref{whist_no_trap}, right panel, which refers to Sim.~9. In this simulation only the lower wave packet is trapped while the upper wave packet is guided outside the soliton. An interpretation of Sim.~9 can be given in terms of geometrical optics. Since the soliton moves along the $x$ axis at a speed $V\sim 0.1$ much smaller than the whistler phase velocity (greater than unity), as a first approximation we neglect the displacement of the soliton. Because of the Doppler shift, the frequency $\omega_0$ measured in the simulation is given by 
\begin{equation}
\omega_0({\bf k},x)=\omega({\bf k},x)+{\bf k\cdot U}(x),
\label{RD_L}
\end{equation}
where $\omega({\bf k},x)$ is the whistler two-fluid dispersion relation in a plasma at rest obtained in the cold limit (see equation~(\ref{RD_w_2fluid}) in Appendix~\ref{two_fl_cold_RD}). The dispersion relation $\omega({\bf k},x)$ is given in terms of $k_y$ and $k_x$ and the density and magnetic field profiles are given by $n=1+n_{sol}$ and ${\bf B}={\bf B}_0+{\bf B}_{sol}$, respectively. In the framework of the geometrical optics, the contours of $\omega_0$ in the plane $(k_x,x)$ for fixed $k_y$ represent the orbits of the whistler wave packet. The solution of the Hamiltonian system 
\begin{equation*}\frac{\partial\omega}{\partial x}=-\dot k_x(t),\quad \frac{\partial\omega}{\partial k_x}=\dot x(t)
\end{equation*} 
gives the evolution of the wave vector and the trajectory of the wave packet. In Fig.~\ref{num_mathematica} we show the contours of $\omega_0$ as defined in equation~(\ref{RD_L}) obtained using the soliton profile of Sim.~9. The wave vector $k_y$ can be estimated from the forcing frequency and injection angle taking into account the doppler shift, giving $k_y\sim1.3$ and $k_y\sim-1.7$. The contours in Fig.~\ref{num_mathematica} show that the orbit corresponding to $\omega_0=3$ is open for the wave packet propagating upward while it is closed for the wave packet propagating downward. A Fourier analysis of the $x$ component of the magnetic field in Sim.~9 confirms that the wave vectors with $|k_y|\sim1.7$ are trapped inside the soliton. Similar results are obtained for Sim.~2 and Sim.~8.

However,  because of the movement along the $x$ axis, the soliton behaves as a \textquotedblleft moving mirror\textquotedblright\ thus causing the  frequency of the injected whistler to change with time. We qualitatively estimate the shifted frequency after the first reflection at the soliton edge $\omega_0^{\prime}=\omega_0-2{\bf k\cdot V}$. The change in frequency could cause the wave to become evanescent. An example is given by Sims.~4 and 5 in a high frequency  whistler regime ($\omega\sim0.1\,\omega_{ce}^{in}$) and using a narrow soliton. In these simulations only the upper wave packet is trapped, as expected, while after the first reflection at the left boundary of the soliton, the lower wave packet becomes evanescent. Fig.~\ref{whist_no_trap}, middle panel, refers to Sim.~4 and shows the contour of $b_x$  and the profiles of the soliton at time $t=30$ (solid line) and $t=0$ (dashed line). In this simulation the lower wave packet has a wave vector $k_y$ estimated to $k_y\sim-5$ (in agreement with the Fourier spectrum of the simulation results). The solution of the Hamiltonian equations for the wave packet with initial conditions $x(0)=0$, $k_x(0)=0$ gives wave packet reflected at nearly $\delta x\sim-1.5$ from the center of the soliton with $k_x\sim25$ (in agreement with the small scales which form in the $x$ direction at the edge of the soliton). In this point the reflected frequency is estimated as $\omega_0^{\prime}\sim\omega_0-2k_x V\sim3.5$, which is below the minimum frequency calculated in correspondence to the edge of the soliton for $k_y\sim-5$, explaining why the lower wave packet  does not propagate after reflection.

\section{Discussion and conclusions}
\label{discussion}

In this paper, we have shown that whistler waves can be efficiently trapped and advected across the magnetic field lines by oblique magnetosonic slow solitons with typical scale length of the order or greater than the ion inertial length. Oblique slow solitons carry a quasi perpendicular density perturbation that is anti-correlated to the magnetic field  perturbation and since the propagation velocity of these solitons is much smaller than the phase velocity of whistlers, they can be viewed by whistlers as quasi stationary inhomogeneities. The soliton plasma density and magnetic field inhomogeneities then act as a true wave guide during whistler propagation. As a result, whistlers can be confined in correspondence to magnetic field depletions associated to density humps (magnetic holes), as we have shown for  frequencies $\omega<\omega_{ce}/2$.  Due to the presence of the magnetic field inhomogeneity, less strict conditions are required for trapping with respect to a channel provided by only a density variation. The possibility to advect whistlers marks an important difference with respect to the case where whistlers have been associated to non propagating mirror modes. Numerical results based on a two-fluid model are in good agreement with theoretical expectations.

Slow solitons, acting as non-linear waves carrier for whistlers,  provide an efficient mechanism to confine whistler energy in space, thus avoiding the spreading of the wave packets, and to transport the whistler energy across magnetic field lines at the soliton typical speed. The model we propose is related to situations often encountered in space plasmas and could explain multi-point Cluster spacecraft observations in the Earth's magnetosphere. The mechanism of whistler trapping that we have discussed relies on the \textquotedblleft inhomogeneous, slow nature\textquotedblright\ of the  wave carrier, that is,   the plasma density and magnetic field strength inhomogeneities are anti-correlated and quasi perpendicular to the background magnetic field and the velocity of propagation is smaller than the whistler phase velocity.  We finally note that other solitonic structures,  propagating slowly with respect to the whistler wave packet, could in principle play the same role in trapping and advecting the whistlers, thus explaining space observations often showing whistler waves in correspondence to local minimum or maximum of the large scale magnetic field and density, respectively.

%

\appendix
\section{Initial conditions for slow magnetosonic solitons}
\label{eq_init_solit}

Below the initial conditions given in the numerical code corresponding to a slow solitary wave~\cite{Ohsawa_Phys_Fluids_1986}. Quantities are normalized to asymptotic equilibrium values outside the soliton.
\begin{equation}
n_{sol} = \frac{A/\alpha}{ \cosh^2   \left[ \sqrt{A/(12\mu)} \, x     \right] },\qquad n = 1 + n_{sol}, \qquad P_{e,i} = P_{0e,i}\, (1+ \Gamma\,n_{sol} )
\end{equation}
\begin{equation}
B_x = \sin{\varphi_0}, \qquad B_y = \cos{\varphi_0} + \left[  \frac{(v^2_{p0} - c^2_s) }{ \cos{\varphi_0}} \right]\,n_{sol}
\end{equation}
\begin{equation}
\begin{split}
B_z = -  \frac{1}{v_{p0}}\frac{(d_e^{-2} - 1)}{d_e^{-2}}\frac{v_{p0}^2(v_{p0}^2-c_s^2)}{(v_{p0}^2 - \sin^2{\varphi_0})} \frac{\sin{\varphi_0}}{\cos{\varphi_0}}\,2\,\sqrt{A/(12\mu)}\,
 \tanh{\left[\sqrt{A/(12\mu)}x\right]}\,n_{sol}
\end{split}
\end{equation}
\begin{equation}
U_x = v_{p0} \, n_{sol}, \qquad U_y = - \left [  \frac{(v_{p0}^2 - c_s^2)}{v_{p0}}    \right] \frac{\sin{\varphi_0}}{\cos{\varphi_0}}
\label{Ux}
\end{equation}
\begin{equation}
\begin{split}
U_z \sim u_{i,z} = -2\,\sqrt{ A/(12\mu)}\,\frac{( v_{p0}^2-c_s^2 )\, (v_{p0}^2\,d_e^2-\sin{\varphi_0}^2)}{\cos{\varphi_0}(v_{p0}^2-\sin{\varphi_0}^2)}\,
 \tanh{\left[\sqrt{A/(12\mu)}x\right]}\,n_{sol}
\label{Uz}
\end{split}
\end{equation}
where $c_s^2=\Gamma(P_{e0}+P_{i0})$ is the sound speed in normalized units and
\begin{equation}
\mu = \frac{v_{p0}\,(v_{p0}^2 - c_s^2)}{4\,d_e^{-2} \, [  v_{p0}^2 - (1 + c_s^2 )/2 ]} 
\left[    1- \frac{(d_e^{-2} -1)^2  \, \sin^2{\varphi_0}}{d_e^{-2} \, (v_{p0}^2 -  \sin^2{\varphi_0})}     \right],
\end{equation}
\begin{equation}
\begin{split}
\alpha = \frac{   3  \, (  v_{p0}^2 - c_s^2 \, \sin^2{\varphi_0}   )  +   (  v_{p0}^2 -  \sin^2{\varphi_0}   ) [  c_s^2 + \Gamma^2 \, (P_{e0} + P_{i0})  ]        }
 {   4 \, v_{p0} \, [v_{p0}^2 - (1+ c_s^2) /2]  },
\end{split}
\end{equation}
\begin{equation}
v_{p0}^2 = \frac{1}{2} \left[ (1 + c_s^2) - \sqrt{(1 + c_s^2)^2 - 4 \,  c_s^2 \, \sin^2{\varphi_0}}     \right].
\end{equation}
\section{Two-fluid cold dispersion relation}
\label{two_fl_cold_RD}

From our two-fluid model, in the cold limit $\omega/k\gg v_{th,e}$ ($v_{th,e}$ is the electron thermal  speed), we get the following dispersion relation for whistler waves propagating in a homogeneous magnetized plasma at rest, at an angle $\theta$ with respect to the equilibrium magnetic field:

\begin{multline}
\label{RD_w_2fluid}
\omega^2 = \frac{1}{2}\frac{v_a^2 k^2}{(1+d_e^2k^2)^2}\left\{ (1+d_e^2k^2) (1+\cos^2{\theta}) + \frac{k^2}{n}\cos^2{\theta}   \right\}\,+ \\
\frac{1}{2}\frac{v_a^2 k^2}{(1+d_e^2k^2)^2}\left\{\sqrt{(1+d_e^2k^2)^2(1-\cos^2{\theta})^2 + 2(1+d_e^2k^2)\frac{k^2}{n}\cos^2{\theta}(1+ \cos^2{\theta}) + \cos^4{\theta}\frac{k^4}{n^2}}           \right\}.
\end{multline}

In equation~(\ref{RD_w_2fluid}) $v_a=B/\sqrt{n}$ is the Alfv\'en velocity in normalized units.

\section*{Acknowledgments}
 
We are pleased to acknowledge the CINECA super computing center (Bologna, Italy) where part of the simulations were performed. A. Tenerani wishes to acknowledge  A. Retin\'o (LPP-CNRS, Observatoire de St.-Maur, France) for the useful discussions.


\begin{thebibliography}{99}

\bibliographystyle{unsrt}


\bibitem{Smith_JGR_1976} E.J. Smith and  B.T. Tsurutani, J. Geophys. Res. {\bf 81}(13), 2261 (1976).

\bibitem{Thorne_nature_1981} R.M. Thorne and B.T. Tsurutani, Nature {\bf 293}, 384 (1981).

\bibitem{Tsurutani_JGR_1982} B.T. Tsurutani and E.J. Smith, J. Geophys. Res. {\bf 87}(A8), 6060 (1982).

\bibitem{Baumjohann_An_Geo_1999} W. Baumjohann, Ann. Geophys. {\bf 17}(12), 1528 (1999).

\bibitem{Dubinin_AG_2007} E.M. Dubinin, Ann. Geophys. {\bf 25}(1), 303 (2007).

\bibitem{Stenzel_JGR_1999} R.L. Stenzel, J. Geophys. Res. {\bf104}(A7), 14379 (1999).

\bibitem{Smith_JGR_1960} E.J. Smith {\it et al.}, J. Geophys. Res. {\bf 65}(3), 815 (1960).

\bibitem{karpman_1981b} V.I. Karpman and R.N. Kaufman, Sov. Phys. JETP {\bf 53}, 956 (1981).

\bibitem{Karpman_Journal_Pl_Phys_1982} V.I. Karpman and R.N. Kaufman, J. Plasma Physics {\bf 27}, 225 (1982).

\bibitem{Streltsov_JGR_2006} A.V. Streltsov {\it et al.}, J. Geophys. Res. {\bf 111}, A03216 (2006).

\bibitem{Angerami_JGR_1970} J.J. Angerami, J. Geophys. Res. {\bf75}(31), 6115 (1970). 

\bibitem{koons_JGR_1989} H.C. Koons, J. Geophys. Res. {\bf94}(A11), 15393 (1989).

\bibitem{Moullard_GRL_2002} O.A. Moullard, Geophys. Res. Lett. {\bf29}(20), 1975 (2002).

\bibitem{stenzel_GRL_1976} R.L. Stenzel, Geophys. Res. Lett. {\bf3}(2), 61 (1976).

\bibitem{McKenzie_PhysPl_2002} J.F. McKenzie and T.B. Doyle, Physics of Plasmas {\bf9}, 55 (2002).

\bibitem{Stasiewicz_PRL_2003} K. Stasiewicz {\it et al.},  Phys. Rev. Lett. {\bf 90}, 085002 (2003).

\bibitem{Stasiewicz_PRL_2004} K. Stasiewicz, Phys. Rev. Lett. {\bf 93}, 125004 (2004).

\bibitem{Stasiewicz_JGR_2005} K. Stasiewicz, J. Geophys. Res. {\bf 110}, A03220 (2005).

\bibitem{Breizman_PRL_2000} B.N. Breizman and A.V. Arefiev, Phys. Rev. Lett. {\bf84}, 3863 (2000).

\bibitem{heading} J. Heading, {\it An Introduction to Phase Integral Methods} (Methuen, 1962)

\bibitem{Ohsawa_Phys_Fluids_1986} Y. Ohsawa, Physics of Fluids {\bf29}(6), 1844 (1986).

\end{thebibliography}
\end{document}